# Abstract Mining

Ellie Small, Javier Cabrera, John B. Kostis, M.D, William Kostis, M.D.

Department of Statistics and Biostatistics, Rutgers University, New Brunswick, NJ

Cardiovascular Institute of New Jersey, Rutgers Robert Wood Johnson Medical School

## Abstract

We have developed an application that will take a "MEDLINE" output from the PubMed database and allows the user to cluster all non-trivial words of the abstracts of the PubMed output. The number of clusters to use can be selected by the user.

A specific cluster may be selected, and the PMIDs and dates for all publications in the selected cluster are displayed underneath. See figure 2, where cluster 12 is selected.

The application also has an "Abstracts" tab, where the abstracts for the selected cluster can be perused. Here, it is also possible to download a HTML file containing the PMID, date, title, and abstract for each publication in the selected cluster.

A third tab is called "Titles", where all the titles for the selected cluster are displayed.

Via a "Use Cluster" button, the selected Cluster can itself be clustered. A "Back" button allows the user to return to any previous state.

Finally, it is also possible to exclude documents whose abstracts contain certain words (see figure 3).

The application will allow researchers to enter general search terms in the PubMed search engine, then use the application to search for publications of special interest within those search terms.

# 1. Introduction

Medical researchers use the PubMed database extensively to search for publications that could help them in their research. In addition, they also search this vast database for publications that will indicate areas that could be of interest for further research.

In the latter case they do not have specific search terms; instead, they use general search terms and look for publications that seem to indicate an unusual link between the subjects of the general search terms and other subjects that could potentially be of interest.

However, those general search terms tend to result in very large numbers of publications, often in the thousands or more. To select the unusual or unexpected ones by reading the abstracts in that case is an extremely time-consuming, at times close to impossible task.

The application we developed aims to solve this problem by accepting all the abstracts for those publications that fulfill the general search terms from the PubMed database, and perform text mining on those abstracts to extract all non-trivial words. The researcher can then repeatedly cluster the publications by commonality of the words in the abstracts in order to find unusual or unexpected combinations of those words. Then upon reading the abstracts for publications with those combinations of words, a determination can be made as to whether the area would be of interest for further research.

# 2. Method

A user who wishes to access this application should first obtain a MEDLINE file from PubMed. Then the user should return to the Abstract Mining application and load the file created; as soon as the file is loaded, the words in its abstracts will be clustered. The user may then change the number of clusters, exclude publications with certain words, cluster the clusters and inspect their contents.

When accessing the Abstract Mining application, a link will appear at the top of the screen: "Go to PubMed" (see figure 1), which may be used to obtain the MEDLINE file needed.  When selected, a new screen/tab will open to the PubMed website. Once in the PubMed website, one or more search terms should be entered. The site will then come back with a number of abstracts. The field "Send to" should be clicked, the radio button "File" should be selected under "Choose Destination", and "Format" should be set to "MEDLINE". A file can then be created by pressing the "Create File" button.

This file does not need to be created via the Abstract Mining application; the user may access the PubMed site independently and create the required file. The option in the Abstract Mining application is provided for convenience only.

In the Abstract Mining application, The MEDLINE file created by the PubMed site may be uploaded via the "Browse" button by selecting the location of the file in the user's file system (usually the downloads directory).

Before uploading the file, a number of clusters may be selected, or the user may leave the default of 6 clusters in place. Note, that at this stage, a maximum of 10 clusters may be selected. If more clusters are required, then the applicable field may be changed to a higher number AFTER the file has been uploaded and initially processed.

Once the MEDLINE file is first processed, the clusters will appear in the section under the "Update" button. The "Cluster" field is initially set to 1, and the PMIDs and dates for cluster 1 will appear underneath the cluster number. The maximum setting for the slider will update to the number of publications in the MEDLINE file minus one. This will allow the user to select as many clusters as possible for the specific MEDLINE file loaded.

The user may change the "Cluster" field in order to see the PMIDs and dates for one of the clusters in the cluster field. The tab "Abstracts" may be selected to see the abstracts for the publications in the selected cluster.

The user may then either select a different number of clusters, or enter words (separated by spaces) in the "Exclude Documents containing these words" field. For either, or both of these changes to take effect, the "Update" button should be selected, which will update the presentation of the clusters underneath it. If words are selected in the "Exclude Documents containing these words" field, all publications with abstracts that contain any of those words will be removed from the selection. See figure 3.

The cluster presentation field underneath the "Update" button has a list of clusters. In parentheses, after each cluster number, the number of publications in that cluster is displayed. Each cluster also displays a list of words; these are the main words that appear in that particular cluster of abstracts.

Once the initial clustering is completed, a user may select one of the clusters (which will automatically update the list of PMIDs and dates as described previously), and select "Use Cluster". This will take all the publications in the selected cluster and cluster them according to the number of clusters requested in the slider and the excluded words. A new cluster presentation will be created that will replace the existing cluster presentation.

This process may be repeated indefinitely. If at any time the user wishes to return to a previous state, the "Back" button may be selected. Repeated use of the "Back" button will eventually return to the original set of publications from the MEDLINE file, minus any excluded publications with words in the "Exclude Documents containing these words" field.

Note that the application has three tabs: "Main", "Abstracts", and "Titles". Everything described so far has been on the "Main" tab.

The "Abstracts" tab will display the PMID, date, title and Abstract of the first publication for the cluster selected on the "Main" tab. The user can scroll through all the publications in the cluster using the "Previous" and "Next" buttons underneath the abstract.

In addition, the "Download Cluster" button may be selected which will create a html file with all PMIDs, dates, titles, and abstracts in the current cluster. See figure 4.

The "Titles" tab will display the PMID, date, and title for all publications in the selected cluster. See figure 5.

Note that this Shiny application is available at https://ellie.shinyapps.io/shiny/.

## 3. Results

Firstly, via the PubMed search engine we selected all publications that contained the phrase "embolic stroke", while excluding the phrase "atrial fibrillation"; there were 10,443 such publications. After loading the abstracts into the application, we selected 23 clusters. At this point the medical researcher was surprised by the occurrence of the word "blood" in combination with the other words in cluster number 8, for which the main words were "cerebral", "stroke", "brain", "blood", "artery", and "ischemic". At the request of the researcher, we then performed a further search of cluster 8, which we itself clustered using another 23 clusters. At this point, the medical researcher determined that cluster 12, which contained the word "progranulin", was of interest. This cluster only contained one publication, "Multiple therapeutic effects of progranulin on experimental acute ischaemic stroke.", and upon checking its abstract the researcher concluded that the link between embolic stroke and progranulin would provide a worthwhile subject for further research.

Secondly, we searched PubMed for the search terms "impedance" and "mismatch", and found 368 publications. We created 20 clusters, and the medical researchers determined that the combination of words in cluster 9, "pressure", "pulmonary", "arterial", "wave", "impedance", and "ventricular", was surprising enough to warrant further investigation. Upon inspection of the 20 publications in this cluster, it was determined that 4 publications in this cluster were of interest for further research; "Systemic vascular hemodynamics and transplanted kidney survival" as it relates to kidneys, "Low compliance rather than high reflection of arterial system decreases stroke volume in arteriosclerosis: a simulation." relating to sepsis, and "Pulmonary impedance and right ventricular-vascular coupling in endotoxin shock." as well as "Aortic and pulmonary input impedance in patients with cor pulmonale.", both of which relate to the lungs.

Finally, we searched PubMed for the search terms "aortic", "stenosis", and "impedance", and found 157 publications, which we clustered into 15 clusters. The word "spacing" in cluster 4 piqued the interest of the medical researchers, and after inspection, both publications in this cluster were determined to be candidates for further research. The publications were "Long-term follow-up impact of dual-chamber pacing on patients with hypertrophic obstructive cardiomyopathy." and "Chronic steroid-eluting lead performance: a comparison of atrial and ventricular pacing."

## 4. Discussion

- We're proposing a new method based on text mining to aid the work of researchers in the medical science.
- Text mining is very helpful for creating a clustering structure into searches in order to allow the researcher to find interesting publications in a very short time, even when the search criteria result in large numbers of publications.
- This method can be expanded to other research oriented websites like Google Scholar, ResearchGate, etc.

# 6. Legend

Figure 1: Initial State

Figure 2: Main page

Figure 3: Excluding Documents

## Abstract Mining

Main | Abstracts | Titles

Go to Pubmed

Upload MEDLINE file from Pubmed: embSnoAF.txt — Upload complete

Exclude documents containing these words: animals rats

How many clusters? 23 — Update

Cluster: 12 — Use Cluster — Back

| PMID | Date |
|---|---|
| 27634955 | 2016-09-17 |
| 20838840 | 2010-09-15 |
| 17898199 | 2007-09-28 |
| 16908573 | 2006-08-16 |
| 16627365 | 2006-04-22 |
| 1903851 | 1991-03-01 |
| 2718208 | 1989-05-01 |
| 2863773 | 1985-06-01 |

```
cluster 1 (1):    mtc spect ecd hmpao brain lesions
cluster 2 (2):    vad og cerebral aorta implantation cfd
cluster 3 (10):   air cerebral embolism artery following blood
cluster 4 (2):    tpa hemorrhage rate combination sm administered
cluster 5 (1):    mice n ang ad ischemic receiving
cluster 6 (3):    cas lesions ischemic plaques plaque cerebral
cluster 7 (6):    mes cerebral kg mg h thrombosis
cluster 8 (164):  cerebral infarction artery stroke patients brain
cluster 9 (4):    mice h cerebral lfus injury rock
cluster 10 (8):   cbf cerebral regions ischemic hours min
cluster 11 (62):  cerebral blood brain ischemic stroke ischemia
cluster 12 (8):   occlusion cerebral min artery aca mca
cluster 13 (2):   perfusion ct brain cerebral images contrast
cluster 14 (7):   cerebral imaging lesion diffusion tavi clinical
cluster 15 (3):   l dilation cerebral lcbf potassium n
cluster 16 (1):   vs use cas cpd cea showers
cluster 17 (1):   angiography mr catheter disease imaging hours
cluster 18 (1):   pa h n mice mrna ischemic
cluster 19 (2):   blood filter suction pericardial flow distal
cluster 20 (30):  cerebral surgery emboli cpb may patients
cluster 21 (2):   brain avms trials clinical one interventional
cluster 22 (12):  tpa cerebral plasminogen activator tissue stroke
cluster 23 (6):   cerebral flow emboli blood arteries r
```

Figure 4: Abstracts tab

## Abstract Mining

Main | Abstracts | Titles

PMID: 25838514  Date: 2015-04-04
Title: Multiple therapeutic effects of progranulin on experimental acute ischaemic stroke.
Abstract:
In the central nervous system, progranulin, a glycoprotein growth factor, plays a crucial role in maintaining physiological functions, and progranulin gene mutations cause TAR DNA-binding protein-43-positive frontotemporal lobar degeneration. Although several studies have reported that progranulin plays a protective role against ischaemic brain injury, little is known about temporal changes in the expression level, cellular localization, and glycosylation status of progranulin after acute focal cerebral ischaemia. In addition, the precise mechanisms by which progranulin exerts protective effects on ischaemic brain injury remains unknown. Furthermore, the therapeutic potential of progranulin against acute focal cerebral ischaemia, including combination treatment with tissue plasminogen activator, remains to be elucidated. In the present study, we aimed to determine temporal changes in the expression and localization of progranulin after ischaemia as well as the therapeutic effects of progranulin on ischaemic brain injury using in vitro and in vivo models. First, we demonstrated a dynamic change in progranulin expression in ischaemic Sprague-Dawley rats, including increased levels of progranulin expression in microglia within the ischaemic core, and increased levels of progranulin expression in viable neurons as well as induction of progranulin expression in endothelial cells within the ischaemic penumbra. We also demonstrated that the fully glycosylated mature secretory isoform of progranulin ( approximately 88 kDa) decreased, whereas the glycosylated immature isoform of progranulin (58-68 kDa) markedly increased at 24 h and 72 h after reperfusion. In vitro experiments using primary cells from C57BL/6 mice revealed that the glycosylated immature isoform was secreted only from the microglia. Second, we demonstrated that progranulin could protect against acute focal cerebral ischaemia by a variety of mechanisms including attenuation of blood-brain barrier disruption, neuroinflammation suppression, and neuroprotection. We found that progranulin could regulate vascular permeability via vascular endothelial growth factor, suppress neuroinflammation after ischaemia via anti-inflammatory interleukin 10 in the microglia, and render neuroprotection in part by inhibition of cytoplasmic redistribution of TAR DNA-binding protein-43 as demonstrated in progranulin knockout mice (C57BL/6 background). Finally, we demonstrated the therapeutic potential of progranulin against acute focal cerebral ischaemia using a rat autologous thrombo-embolic model with delayed tissue plasminogen activator treatment. Intravenously administered recombinant progranulin reduced cerebral infarct and oedema, suppressed haemorrhagic transformation, and improved motor outcomes (P = 0.007, 0.038, 0.007 and 0.004, respectively). In conclusion, progranulin may be a novel therapeutic target that provides vascular protection, anti-neuroinflammation, and neuroprotection related in part to vascular endothelial growth factor, interleukin 10, and TAR DNA-binding protein-43, respectively.

Previous | Next

Download Cluster

Figure 5: Titles tab

## Abstract Mining

Main | Abstracts | Titles

| PMID | Date | Title |
| --- | --- | --- |
| 29387392 | 2018-02-02 | Effect of recombinant human prourokinase on thrombolysis in a rabbit model of thromboembolic stroke. |
| 28035007 | 2016-12-31 | Inhibition of Factor XIa Reduces the Frequency of Cerebral Microembolic Signals Derived from Carotid Arterial Thrombosis in Rabbits. |
| 27262051 | 2016-06-05 | A rabbit model of cerebral microembolic signals for translational research: preclinical validation for aspirin and clopidogrel. |
| 26763925 | 2016-01-15 | Cerebrolysin dose-dependently improves neurological outcome in rats after acute stroke: A prospective, randomized, blinded, and placebo-controlled study. |
| 25060418 | 2014-07-26 | A dose-response study of thymosin beta4 for the treatment of acute stroke. |
| 17588313 | 2007-06-26 | Effects of microplasmin on recovery in a rat embolic stroke model. |
| 10699454 | 2000-03-04 | Enhanced neuroprotection and reduced hemorrhagic incidence in focal cerebral ischemia of rat by low dose combination therapy of urokinase and topiramate. |
| 8842419 | 1996-08-01 | Neuroprotective properties of a protein kinase inhibitor against ischaemia-induced neuronal damage in rats and gerbils. |
| 8832673 | 1996-08-01 | The effect of oral antiplatelet agents on tissue plasminogen activator-mediated thrombolysis in a rabbit model of thromboembolic stroke. |
| 7495108 | 1995-07-01 | Beneficial effect of CV-4151 (Isbogrel), a thromboxane A2 synthase inhibitor, in a rat middle cerebral artery thrombosis model. |
| 8327988 | 1993-05-01 | Effect of Y-20811, a thromboxane synthetase inhibitor, on photochemically induced cerebral embolism in rabbits. |
| 2523649 | 1989-04-17 | Cerebrovascular, biochemical, and cytoprotective effects of isradipine in laboratory animals. |